# Interaction of ion-acoustic solitons with electron beam in warm plasmas with superthermal electrons

A R Esfandyari-Kalejahi and E Saberian

Department of Physics, Faculty of Sciences, Azarbaijan University of Tarbiat Moallem, 53714-161, Tabriz, Iran
(Dated: 18 Apr. 2012)

E-mail: s.esaberian@azaruniv.edu and esaberyan@yahoo.com

**Abstract**
Propagation of ion-acoustic solitary waves (IASWs) is studied using the hydrodynamic equations coupled with the Poisson equation in a warm plasma consisting of adiabatic ions and superthermal (Kappa distributed) electrons in presence of an electron-beam component. In the linear limit, the dispersion relation for ion-acoustic (IA) waves is obtained by linearizing of basic equations. On the other hand, in the nonlinear analysis, an energy-balance like equation involving Sagdeev's pseudo-potential is derived in order to investigate arbitrary amplitude IA solitons. The Mach number range is determined in which, propagation and characteristics of IA solitons are analyzed both parametrically and numerically. The variation of amplitude and width of electrostatic (ES) excitations as a result of superthermality (via $\kappa$) and also the physical parameters (ion temperature, soliton speed, electron-beam density and electron-beam velocity) are examined. A typical interaction between IASWs and the electron-beam in plasma is confirmed. It is also found that ES solitons with both negative polarity (rarefactive solitons) and positive polarity (compressive solitons) can coexist in plasma. Furthermore, the small but finite amplitude limit of IASWs is investigated and it is showed that this limit is restricted to negative potential pulses.

Keywords: Suprthermal electrons, Kappa distribution, Ion-acoustic solitary waves, Sagdeev's pseudo-potential

PACS numbers: 52.35.Ex, 52.35.Mw, 52.35.Fp, 52.35.Sb, 52.40.Mj, 94.05.Fg

## 1. Introduction

Nowadays, nonlinear propagation of the electrostatic (ES) disturbances in in space and laboratory plasmas has received a considerable attention and been extensively studied [1-8]. There are many records of satellite observations that indicate the existence of nonlinear solitary ES structures in space plasmas [9-11]. It is believed that studying the ion-acoustic (IA) modes is very useful in understanding the low frequency component of broadband ES excitations observed in different regions of the earth's magnetosphere and the other plasma environments. So, numerous investigations about nonlinear ES modes have been devoted to the studying of various aspects of ion-acoustic solitary waves (IASWs) in multi-component plasmas for many years [12-19]. The small-amplitude nonlinear waves in plasma generically can be described either by the Korteweg-de Veries (KdV) equation or nonlinear Schrödinger equation (NLSE), derived via reductive perturbation techniques, but the nonlinear waves with arbitrary amplitude can be analyzed using the standard Sagdeev's pseudopotetial approaches [20].

    Of particular interest is the case when an electron beam is present in the plasma. Such a situation is typically encountered in the upper layers of the magnetosphere, where the

coexistence of two different electron populations (say, cold, i.e., "inertial" and warm, i.e., "energetic" ones) have been reported by satellite missions, e.g. the FAST at the auroral region [21,22], the S3-3[23], Viking[24], GEOTAIL and POLAR[22,25] missions. Such a beam-plasma system has also been created in laboratory [26–28] where modulated envelope wave packets were created and observed. From a theoretical point of view, the existence of an electron beam in the plasma has been shown to modify the properties and conditions for existence of arbitrary amplitude localized ES nonlinear excitations [29–31], in addition to small amplitude solitary pulses [32–34]. Such theoretical considerations have later been extended to the (higher-frequency) electron-acoustic (EA) waves [35] (related to Broadband Electrostatic Noise, BEN, in the Earth's auroral region), supported by numerical simulations [36] and also to ion-beam effects [37]. In order to explain the experimental results, the effects of various plasma parameters on structure of small and large amplitude ES solitons have been studied by many authors in electron beam-plasma [30,38-40]. Considering a plasma system comprised of adiabatic warm ions, nonisothermal electrons and weakly relativistic electron beam has been yielded a considerable modification on structure of IASWs [41]. Recently, arbitrary amplitude EA solitons in the presence of an electron beam and superthermal electrons has been studied [42].

It is confirmed that particle velocity distribution functions (VDFs) in space plasmas as well as in many experimental plasmas often show non-Maxwellian tails decreasing as a power law of the velocity. Nonthermal particle distributions are ubiquitous at high altitudes in the solar wind and many space plasmas, their presence being widely confirmed by spacecraft measurements [43-47]. Such deviations from the Maxwellian distributions are also expected to exist in any low-density plasma in the Universe, where binary collisions of charges are sufficiently rare. The suprathermal populations are well described by the so-called Kappa velocity distribution function, as shown for the first time by Vasyliunas in 1968 [48]. Such distributions have high energy tails deviated from a Maxwellian and decreasing as a power law in particle speed:

$$f_i^\kappa(r,v) = \frac{n_i}{2\pi(\kappa w_{\kappa i}^2)^{3/2}} \frac{\Gamma(\kappa+1)}{\Gamma(\kappa-1/2)\Gamma(3/2)} \left(1 + \frac{v^2}{\kappa w_{\kappa i}^2}\right)^{-(\kappa+1)}, \qquad (1)$$

where, $w_{ik} = \sqrt{(2\kappa-3)kT_i/\kappa m_i}$ is the thermal velocity, $m_i$ the mass of the particles of species $i$, $n$ their number density, $T$ their equivalent temperature, $v$ the velocity of the particles, and $\Gamma(x)$ is the Gamma function. The spectral index must take sufficiently large values $\kappa > 3/2$ to keep away from the critical value $\kappa_c = 3/2$, where the distribution function (1) collapses and the equivalent temperature is not defined. The value of the index $\kappa$ determines the slope of the energy spectrum of the suprathermal particles forming the tail of the velocity distribution function. Using the Kappa distribution function and integrating over velocity space, one can obtain the number density of the corresponding plasma species. In the $\kappa \to \infty$ limit, the Kappa function degenerates into a Maxwellian. Note also that different mathematical definitions of Kappa distributions are commonly used and various authors characterize the power law nature of suprathermal tails in different ways.

Distributions with suprathermal tails have been observed in various space plasmas [49-55] and various mechanisms have been proposed to explain the origin of the suprathermal tails of the VDFs and occurrence of Kappa-like distributions in the solar wind, the corona and other space plasmas. The first one was suggested by Hasegawa et al. in 1985 [56] who showed that when the plasma immersed in a suprathermal radiation field suffers velocity-space diffusion which is enhanced by the photon-induced Coulomb-field fluctuations. This enhanced diffusion universally produces a power-law distribution. Collier in 1993 uses random walk jumps in velocity space whose path lengths are governed by a power or L´evy flight probability distribution to generate Kappa-like distribution functions [57]. Treumann developed a kinetic theory to show that Kappa-like VDFs correspond to a particular thermodynamic equilibrium state [58]. Leubner in 2002 [59] shows that Kappa-like distributions can result as a consequence of the entropy generalization in nonextensive Tsallis statistics [60], physically related to the long range nature of the Coulomb potential, turbulence and intermittency [61,62]. The Kappa distribution is equivalent to the $q$ distribution function obtained from the maximization of the Tsallis entropy. Systems subject to long-range interactions and correlations are fundamentally related to non-Maxwellian distributions [63]. Livadiotis and McComas in 2009 also examined how Kappa distributions arise naturally from Tsallis statistical mechanics and provide a solid theoretical basis for describing complex systems [64]. The generation of suprathermal electrons by resonant interaction with whistler waves in the solar corona and wind was suggested by Vocks and Mann in 2003 [65] and by Vocks et al. in 2008 [66].

The superthermal particles have important consequences concerning the acceleration and the temperature that are well evidenced by the kinetic approach where no closure requires the distributions to be nearly Maxwellians. Moreover, the presence of superthermal particles takes an important role in the wave-particle interactions. In many circumstances, the wave-particle interactions can be made responsible for establishing non-Maxwellian particle distribution functions with an enhanced high energy tail and shoulder in the profile of the distribution function [67]. In turn, the general plasma dynamics and dispersion properties are also altered by the presence of nonthermal populations. Whatever the mechanisms of suprathermal tails formation, the kappa function is a useful mathematical tool to generalize the velocity distributions to the observed power law functions, specially the particular Maxwellian VDF corresponding to the specific value of $\kappa \to \infty$.

Several studies have been done connected to various aspects of ES excitations in plasmas using Kappa distribution function for the electron component [42,68-73]. However, most of the theoretical studies on IASWs in superthermal plasmas considered cold ion-component in their model [69-72].

In most of the astrophysical and experimental plasmas, the ions include the thermal effects and so the formation of the solitary waves is sensitive to the ion temperature. Anyway, the warm ion density has a transcendental form while solving the adiabatic fluid equations and hence the ion temperature effect essentially modifies the characteristics of the IA excitations. However, we have to notice that the existence of solitary waves depends on finite ion-temperature effect. So far, some researchers have taken into account the ion thermal effects on the features of the IASWs in typical plasmas and have showed a more realistic picture for usual plasmas [74-80].

The aim of this paper is to study the effects of various parameters on propagation of IASWs in presence of an electron beam in a superthermal plasma and to show a typical interaction between electron beam and IA solitons. In this paper, we shall follow the work of Saini and Kourakis [72] for investigation of the IA solitons interaction and electron beem-plasma in a warm plasma consisting of adiabatic ions, an electron-beam and superthermal electrons. First, in the linear analysis, the dispersion relation will be derived here. Further, in the nonlinear case, we shall employ the standard pseudo-potential theory for investigation of fully nonlinear IASWs and their characteristics. The energy integral and energy-balance equation will be derived and the small-amplitude limit will be discussed. Then, the Mach number range will be analyzed and also soliton structures will be discussed therein.

The paper is organized as follows: in Section 2, the basic fluid equations for study of linear and nonlinear IA waves are presented. Section 3 contains a brief discussion about linear behavior of IA modes including the linear dispersion relation. In Section 4, an energy-balance equation is derived together with the Sagdeev's pseudo-potential function for analysis of nonlinear IASWs. Section 5 is devoted to small-amplitude IA solitons. In Section 6, we shall apply the conditions for existence of arbitary amplitude IASWs and investigate parametrically the Mach number range for existence of solitary waves. These are shown graphically with variation of different parameters. The structure of solitary waves and effects of various parameters, e.g. superthermality index, ion temperature, soliton speed and electron beam parameters (density and velocity), on soliton amplitude and steepness will be discussed numerically in section 7. Finally, Section 8 is devoted to concluding remarks.

**2. Theory and the model**

We consider a collisionless and unmagnetized plasma consisting of warm ions, electron beam and superthermal electrons modeled by a kappa distribution. In presence of an electron beam-plasma the usual plasma wave modes, e.g., ion-acoustic, ion-cyclotron, etc., are modified because of interaction of plasma waves with electron beam component. For these plasma modes, the relatively massive ions do not participate in the wave dynamics and are treated as an immobile charge background distribution.

So, the propagation of IA waves in beam-plasma background is governed by a system of fluid equations for the ion fluid, distinguished by using the index "$i$", and fluid equations for the electron beam, distinguished by using the index "$b$", all of them are coupled through the Poisson equation as

$$\frac{\partial n_i}{\partial t} + \frac{\partial}{\partial x}(n_i v_i) = 0, \quad (2)$$

$$\frac{\partial v_i}{\partial t} + v_i \frac{\partial v_i}{\partial x} = -\frac{Ze}{m_i}\frac{\partial \phi}{\partial x} - \frac{1}{n_i m_i}\frac{\partial p_i}{\partial x}, \quad (3)$$

$$\frac{\partial n_b}{\partial t} + \frac{\partial}{\partial x}(n_b v_b) = 0, \quad (4)$$

$$\frac{\partial v_b}{\partial t} + v_b \frac{\partial v_b}{\partial x} = \frac{e}{m_e}\frac{\partial \phi}{\partial x}, \quad (5)$$

$$\frac{\partial^2 \phi}{\partial x^2} = -4\pi e(n_i Z - n_e - n_b), \quad (6)$$

where $n_i$, $n_b$, $n_e$, $v_i$ and $v_b$ refer to the number density of the ions, number density of beam, number density of the electrons, ion fluid velocity and beam fluid velocity, respectively. Furthermore, $m_i$ is the mass of the ions, $m_e$ is the mass of the electrons, $e$ is the magnitude of electron charge, $Z$ is the ion charge number, $p_i$ is the ion pressure and $\phi$ is the electrostatic potential.

The right-hand side of equation (6) cancels at equilibrium, due to the overall quasi-neutrality condition, implies

$$Z = \alpha + \nu, \quad (7)$$

where $\alpha = \frac{n_{e0}}{n_{i0}}$ and $\nu = \frac{n_{b0}}{n_{i0}}$ refer to the ratio of unperturbed electron to ion densities and the ratio of unperturbed beam to ion densities, respectively.

The electron number density is assumed to be Kappa distributed, namely

$$n_e = n_{e0}\left[1 - \frac{e\phi}{(\kappa-\frac{3}{2})k_B T_e}\right]^{-\kappa+\frac{1}{2}}, \quad (8)$$

where the real parameter $\kappa$ measures the deviation from the Maxwell-Boltzmann equilibrium (which is recovered in the limit $\kappa \to \infty$). It is mentioned that spectral index must take sufficiently large values $\kappa > 3/2$ in order for a physically meaningful thermal speed to be defined.

Equations (2)-(6) may be cast in a reduced form, for convenience. We Scale the time $t$ and distance $x$ by the inverse of ion plasma frequency $\omega_{pi}^{-1} = (\frac{m_i}{4\pi n_{i0} Z^2 e^2})^{1/2}$ and the Debye length $\lambda_{De} = \left(\frac{k_B T_e}{4\pi Z n_{i0} e^2}\right)^{\frac{1}{2}}$, respectively. Furthermore, we scale the number densities of ion $n_i$ and beam $n_b$ by the unperturbed ion density $n_{i0}$, ion fluid velocity $v_i$ and beam fluid velocity $v_b$ by the ion-sound speed $c_s = (\frac{Zk_BT_e}{m_i})^{1/2}$, the ion pressure $p_i$ by $p_{i0} = n_{i0}k_BT_i$ and finally the electrostatic potential $\phi$ by the $k_BT_e/e$.

The equations (2)-(6) can thus be rewritten into the normalized equations as follows

$$\frac{\partial n_i}{\partial t} + \frac{\partial}{\partial x}(n_i v_i) = 0, \tag{9}$$

$$\frac{\partial v_i}{\partial t} + v_i \frac{\partial v_i}{\partial x} = -\frac{\partial \phi}{\partial x} - \frac{\gamma \sigma}{Z} n_i^{\gamma-2} \frac{\partial n_i}{\partial x}, \tag{10}$$

$$\frac{\partial n_b}{\partial t} + \frac{\partial}{\partial x}(n_b v_b) = 0, \tag{11}$$

$$\frac{\partial v_b}{\partial t} + v_b \frac{\partial v_b}{\partial x} = \frac{1}{\mu Z}\frac{\partial \phi}{\partial x}, \tag{12}$$

$$\frac{\partial^2 \phi}{\partial x^2} = -n + \frac{Z-\nu}{Z}\left(1 - \frac{\phi}{\kappa - \frac{3}{2}}\right)^{-\kappa + \frac{1}{2}} + \frac{n_b}{Z}, \tag{13}$$

where $\sigma = \frac{T_i}{T_e}$ and $\mu = \frac{m_e}{m_i}$ ($\simeq \frac{1}{1836}$) are the fractional ion to electron temperature and ratio of electron mass to ion mass, respectively. Here, $\gamma$ is polytropic index and taken in the following to be $\gamma = 3$ for adiabatic ions.

## 3. Linear dispersion relation

To elucidate the basic physics of the IA modes we need to obtain the linear dispersion relation for these ES modes. Neglecting the nonlinear terms in equations (8)-(12) and combining the Fourier transform of the resultant equations, by assuming that all the perturbed quantities vary as $\exp[i(kx - \omega t)]$, the linear dispersion relation is obtained as

$$1 + \frac{(1-\nu/Z)}{k^2}\left(\frac{\kappa-1/2}{\kappa-3/2}\right) - \frac{1}{(\omega^2 - 3\frac{\sigma}{Z}k^2)} - \frac{\nu}{\mu Z}\frac{1}{(\omega - kU_0)^2} = 0, \tag{14}$$

where $\omega$ and $k$ are the wave frequency and the wave number, respectively, and $U_0 = u_{b0}/c_s$ is the reduced beam velocity.

In deriving equation (14), we restrict the beam density and the beam velocity to very low values, i.e. by taking $\nu U_0 \ll 1$, which correspond to a finite but negligible charge current, $j = n_{b0}u_{b0} \simeq 0$, at equilibrium. This restriction is necessary in order for the ES character of excitations to be preserved.

In linearized dispersion relation given in equation (14), the second term contains superthermality effect of electrons via $\kappa$, the third term contains thermal effects of ions via $\sigma$ and finally the fourth one includes effect of beam drift via $U_0$ on dispersion of IA waves.

Figure 1 depicts behavior of the linear dispersion relation for IA modes obtained in equation (14), for various values of $\kappa$ to show superthermality effect on these modes. The solid curve corresponds to $\kappa = 16$ that exhibit pseudo-Maxwellian limit (high value of $\kappa$ or less superthermality) and the other curves show variations from Maxwellian limit (small value of $\kappa$ or increasing of superthermality). One can see that with increasing the superthermality, the slope of $\omega$ versus $\kappa$ decreases and so the phase speed $v_{ph} = \frac{\omega}{k}$ for IA waves slow down to less values, suggesting that IA modes have an upper boundary phase speed at Maxwellian limit.

## 4. Nonlinear analysis

Anticipating a stationary profile solitary wave solution, we make all the dependent variables depend on a single variable $\xi = x - Mt$, where $M$ is the Mach number (the velocity of the solitary wave normalized to the ion-sound speed $c_s$). Imposing appropriate boundary conditions for localized waves: $\phi \to 0$, $n_i \to 1$, $v_i \to 0$, $n_b \to \nu$, $u_b \to U_0$ as $\xi \to \pm\infty$, we can integrate equations (9) and (10) to obtain

$$n_i = \frac{M}{M - v_i}, \tag{15}$$

$$\frac{3\sigma}{Z}n_i^4 - \left(M^2 - 2\phi + 3\frac{\sigma}{Z}\right)n_i^2 + M^2 = 0, \tag{16}$$

where the resultant density equation (15) has been combined with the fluid velocity equation in order to eliminate the velocity $v_i$ in the latter, thus equation (16) has been obtained.

*Cold-ion limit.* Let us briefly consider, for later reference, the cold-ion limit. Setting $\sigma = 0$ in equation (16), one finds the cold-ion solution as $n_i = \frac{|M|}{(M^2 - 2\phi)^{1/2}}$, which the reality condition $M^2 \geq 2\phi$ has to be imposed here. This requirement restricts for positive potential structures only, in which $\phi \leq \frac{M^2}{2}$, where the reality condition for ion density is satisfied. Also notice that $n_i \to 1$ for $\phi \to 0$, as expected.

From equation (16), two solutions for the density $n_i$ are obtained as

$$n_{i\pm} = \frac{1}{\sqrt{6\sigma/Z}}\left\{M^2 + \frac{3\sigma}{Z} - 2\phi \pm \sqrt{\left(M^2 + \frac{3\sigma}{Z} - 2\phi\right)^2 - \frac{12\sigma}{Z}M^2}\right\}^{1/2}. \tag{17}$$

The distinction among the two solutions imposes an investigation in terms of the interplay among the values of parameters $M$, $\sigma$ and $\phi$. Three criteria need to be considered here
(i) First, the reality of $n_i$ needs to be ensured. Furthermore,
(ii) One has to impose $n_i \to 1$ as $\phi \to 0$, i.e. the equilibrium state must be accessible as expected. Finally,
(iii) $n_i$ should be analytical (non-singular) for all non-negative values of $\sigma \geq 0$, and in fact tend smoothly to the correct cold-ion limit $\frac{|M|}{(M^2 - 2\phi)^{1/2}}$ for $\sigma \to 0$ (recall the cold-ion limit above). The analysis, however tedious, is quite straightforward. We find that (only) $n_{i-}$ satisfies these criteria, if

(a) $|M| > \sqrt{3\sigma/Z}$ (necessary for $n_{i-} \to 1$ as $\phi \to 0$) and

(b) $\phi \leq \frac{\left(M - \sqrt{3\sigma/Z}\right)^2}{2} \equiv \phi_m^+$, \tag{18}

which is necessary because $n_{i-}$ must be real. We see that this requirement satisfied only by positive potential structures and positive solutions for $\phi$ are limited to $\phi \leq \frac{\left(M - \sqrt{3\sigma/Z}\right)^2}{2}$. Note that the cold-ion limit is smoothly recovered upon setting $\sigma \to 0$. On the other hand, we find that $n_{i+} \to 1$ as $\phi \to 0$ if the inequality $|M| < \sqrt{3\sigma/Z}$ is satisfied. However, since $n_{i+}$ isn't analytical for $\sigma \to 0$ we shall therefore abandon this solution.

After integrating equations (11) and (12) for the electron beam, we find

$$n_b = v\frac{M - U_0}{M - u_b} \tag{19}$$

$$\frac{1}{2}u_b^2 - Mu_b + MU_0 - \frac{U_0^2}{2} - \frac{\phi}{Z\mu} = 0. \tag{20}$$

Combining equations (19) and (20), we find the electron beam density as

$$n_b = \frac{v}{\sqrt{1 + \frac{2\phi}{Z\mu(M - U_0)^2}}}. \tag{21}$$

For the reality of electron beam density $n_b$ in equation (21), the following inequality needs to be satisfied

$$\phi \geq -\frac{Z\mu(M - U_0)^2}{2} \equiv \phi_m^-. \tag{22}$$

We find immediately that this solution satisfied only by negative potential structures and upper amplitude of $\phi$ (in fact in absolute value) restricted to values of $|\phi| \leq \frac{Z\mu(M - U_0)^2}{2}$.

From this inequality, we expect that negative potential structures occur in small values (because of smallness of $\mu$).

It is deduced that from inequalities (18) and (22), that potential structures with both positive polarity and negative polarity are permitted in plasma and allowed domain for potential is restricted to $\phi_m^- \leq \phi \leq \phi_m^+$, where the critical potential $\phi_m^-$ and $\phi_m^+$ were defined above and $\phi_m^- \leq 0 \leq \phi_m^+$. Furthermore, it is mentioned that existence regions of positive potential structures are specified by the ions, while negative ones are characterized by the beam.

Inspired by Ref. 81, we shall now introduce the ansatz:
$$\theta = exp[\cosh^{-1}(\chi)], \tag{23}$$
where $\chi = \frac{(M^2 + \frac{3\sigma}{Z} - 2\phi)}{|M|\sqrt{12\sigma/Z}}$.

The expression given in equation (17) for the ion density may now be cast in the form
$$n_{i_\pm} = \frac{|M|^{1/2}}{(3\sigma/Z)^{1/4}} \theta^{\pm \frac{1}{2}}. \tag{24}$$

Note that the inverse hyperbolic cosine function is defined for values of the argument larger than zero ($\chi > 0$), indeed, this requirement is ensured by the reality of $n_{i_\pm}$ given in Eq. (17).

Substituting $n_b$ and $n_{i_-}$ from equations (21) and (24) into the Poisson equation (13), multiplying the resulting equation by $\frac{d\phi}{d\xi}$, integrating and applying the boundary conditions, $\phi \to 0$ and $\frac{d\phi}{d\xi} \to 0$ at $\xi \to \pm\infty$, leads to following energy-balance equation
$$\frac{1}{2}\left(\frac{d\phi}{d\xi}\right)^2 + \psi(\phi, M) = 0, \tag{25}$$
where $\psi(\phi, M)$ is a Sagdeev-type function modeling a pseudo-potential in which a fictitious particle of unit mass, with position $\phi$ and velocity $\frac{d\phi}{d\xi}$ oscillates. The pseudo-potential $\psi(\phi, M)$ is given by
$$\psi(\phi, M) = -\frac{1}{3}\left(\frac{3\sigma}{Z}|M|^6\right)^{\frac{1}{4}}\left(\theta^{-\frac{3}{2}} - \theta_0^{-\frac{3}{2}}\right) - \left(\frac{3\sigma}{Z}|M|^6\right)^{\frac{1}{4}}\left(\theta^{\frac{1}{2}} - \theta_0^{\frac{1}{2}}\right)$$
$$+ \frac{Z-\nu}{Z}\left[1 - \left(1 - \frac{\phi}{\kappa - \frac{3}{2}}\right)^{-\kappa + \frac{3}{2}}\right] + \nu\mu(M-U_0)^2\left[1 - \left(1 + \frac{2\phi}{Z\mu(M-U_0)^2}\right)^{\frac{1}{2}}\right]. \tag{26}$$

Here, we have used the notation $\theta_0 = \theta(\phi = 0)$ for the unperturbed quantities.

## 5. Small amplitude theory

Let us consider the small-amplitude limit in the above analysis. Expanding the potential $\psi(\phi, M)$ in equation (26) near $\phi = 0$, we obtain
$$\psi(\phi, M) \approx \frac{\psi_0''}{2}\phi^2 + \frac{\psi_0'''}{6}\phi^3, \tag{27}$$
where $\psi_0'' = \frac{\partial^2}{\partial\phi^2}\psi\Big|_{\phi=0}$ and $\psi_0''' = \frac{\partial^3}{\partial\phi^3}\psi\Big|_{\phi=0}$ are computed from equation (26) as
$$\psi_0'' = \frac{1}{M^2 - \frac{3\sigma}{Z}} - \frac{Z-\nu}{Z}\left(\frac{2\kappa-1}{2\kappa-3}\right) + \frac{\nu}{\mu Z^2(M-U_0)^2}, \tag{28}$$
$$\psi_0''' = \frac{1}{3\sigma/Z}\left[\frac{-1}{M^2 - 3\sigma/Z} + \frac{(M^2 + 3\sigma/Z)^2}{(M^2 - 3\sigma/Z)^3}\right] - \frac{1}{(M^2 - 3\sigma/Z)^2} + \frac{Z-\nu}{Z}\frac{(2\kappa-1)(2\kappa+1)}{(2\kappa-3)^2} - \frac{3\nu}{\mu^2 Z^3(M-U_0)^4}. \tag{29}$$

Inserting equations (27) into equation (25) and integrating, we obtain (provided that $\psi_0'' < 0$) a solitary solution in the form
$$\phi(\xi) = \frac{-3\psi_0''}{\psi_0'''} sech^2\left(\frac{\sqrt{-\psi_0''}}{2}\xi\right). \tag{30}$$

The pulse profile given by equation (30) is identical to the soliton solution of the KdV equation, which obtains by use of the reductive perturbation method, for example see Ref. 82, page 260.

It is important to notice that the soliton width $L = 2/\sqrt{-\psi_0''}$ and amplitude $\phi_m = -\frac{3\psi_0''}{\psi_0'''}$ satisfy $\phi_m L^2 = \frac{12}{\psi_0'''} = constant$, as known from the KdV theory.

## 6. Arbitrary amplitude analysis: parametric investigation

For IASW solution of equation (25) to exist, the following requirements for the Sagdeev potential must be satisfied:

(i) $\psi(\phi = 0, M) = d\psi(\phi, M)/d\phi|_{\phi=0} = 0$, which indicates that both the electric field and the charge density be zero far from the localized potential solitary structures,

(ii) $d^2\psi(\phi, M)/d\phi^2|_{\phi=0} < 0$, so that the fixed point at the origin is unstable i.e. $\psi(\phi, M)$ has a local maximum at the origin, and finally,

(iii) $\psi(\phi, M) < 0$ in the region $0 < |\phi| < |\phi_m|$; which $\phi_m > 0$ ($\phi_m < 0$) indicates the maximum (minimum) critical value of $\phi$ for positive (negative) potential well at which compressive (rarefactive) solitary waves exist.

### 6.1. Soliton speed threshold $M_1$

The roots of $d^2\psi(\phi, M)/d\phi^2|_{\phi=0} = 0$ in terms of Mach number defines the lower limit of the Mach number (Mach number threshold $M_1$), after which solitary wave can propagate in plasma. Imposing condition (ii) to equation (26) implies that the following inequality should hold

$$\frac{1}{M^2 - \frac{3\sigma}{Z}} - \frac{Z-\nu}{Z}\left(\frac{2\kappa-1}{2\kappa-3}\right) + \frac{\nu}{\mu Z^2 (M-U_0)^2} \leq 0, \quad (31)$$

where the equality holds for $M = M_1$. Strictly speaking, this relation delimits the regions of possible values of the excitation speed $M$ (Mach number), at which localized pulses may propagate.

It is constructive to discuss some boundary conditions in equation (31), in order to investigate the various parameters effect and also recover the expected results in previous works.

*Maxwellian cold e-i plasma limit.* Setting $\nu \to 0$, $\sigma \to 0$ and $\kappa \to \infty$ in equation (31), the known Mach number condition $M \geq 1$ is obtained (in which $M_1 = 1$) that contain supersonic nature of IA solitary waves in ordinary *e-i* plasmas [20].

*Cold-ion limit.* Setting $\sigma \to 0$ in equation (31), we obtain the Mach number threshold condition in the case of zero temperature ions as

$$\frac{1}{M^2} - \frac{Z-\nu}{Z}\left(\frac{2\kappa-1}{2\kappa-3}\right) + \frac{\nu}{\mu Z^2 (M-U_0)^2} \leq 0. \quad (32)$$

This case has been investigated, in details, by Saini and Kourakis [72]. Two important results in this analysis, which are hold in our research, are the lower the sound speed ($M_1$) by the superthermality effect and to increase in threshold Mach number by the beam effect. However, the combined effect of the electron beam and the superthermality of the background consist of a competition among them.

*Ion temperature effect (no beam and no superthermality).* By setting $\nu \to 0$ and $\kappa \to \infty$ in equation (31), we can derive the effect of the ion temperature on Mach number threshold for IA solitary waves in a Maxwellian *e-i* plasma as $M_1^{(\sigma)} = (1 + 3\sigma/Z)^{\frac{1}{2}}$. Note that $M_1^{(\sigma)} > 1$, suggesting that the effect of the ion temperature is to increase the threshold for electrostatic excitations to exist. This limit is compatible with the work of Nejoh by imposing no positron population in plasma [13].

*Ion temperature and superthermality effect (no beam).* Similar to previous limits, in absence of electron beam ($\nu \to 0$) we obtain threshold Mach number as

$$M_1^{(\sigma,\kappa)} = \left(\frac{3\sigma}{Z} + \frac{2\kappa-3}{2\kappa-1}\right)^{\frac{1}{2}}. \tag{33}$$

In this case, we see again that there is a competition between the ion temperature effect and superthermality, which tends to increase and decrease, respectively, the soliton speed threshold. Note that depend on ion temperature and superthermality, both subsonic and supersonic solitons can propagate in plasma.

*Universal analysis of $M_1$*
For completeness of this section, we solve equation (31) numerically and analyze the variation of the soliton speed threshold $M_1$ with respect to each parameter. Solving equation (31) yields four roots; anyway only real positive roots of $M_1$ are acceptable. Among two positive roots of $M_1$, only one of them is compatible with the necessary condition $|M| > \sqrt{3\sigma/Z}$ for realistic values of physical parameters and so it is physically meaningful.

Without loss of generality, we consider $Z = 1$ for this study. The variation of the threshold soliton speed $M_1$ with respect to the ion temperature ($\sigma$), the electron-beam density ($\nu$) and the beam velocity ($U_0$) for fixed values of $\kappa$ have displayed in figures 2. In these figures, the solid curve represents the pseudo-Maxwellian limit (high $\kappa$) and the other curves display deviations from the Maxwellian limit via superthermality effect (note that the values of $\kappa$ above ~10 is practically tantamount to Maxwellian). From these figures, it is seen that $M_1$ increases with an increase in ion temperature, electron-beam density and beam velocity (for positive $U_0$) and also with an increase in the value of $\kappa$. Knowing that a small value of $\kappa$ corresponds to more superthermal particles, so an increase in superthemality lowers the threshold Mach number $M_1$.

In figure 2(c), we have considered both positive and negative values of the beam velocity $U_0$, showing both co-propagation and counter-propagation of ES solitary waves and beam flow. In the positive regions (co-propagation case), $M_1$ increases with an increase in beam velocity, but in the negative regions (counter-propagation case) the result is reversed and an increase in the absolute value of beam velocity leads to a decrease in $M_1$.

In general, the presence of warm ions and electron beam cause the soliton speed threshold $M_1$ to increase in comparison with ordinary cold *e-i* plasmas, but the presence of superthermal electrons results in the soliton speed threshold $M_1$ to lower as compared to Maxwellian plasmas.

*6.2. Upper soliton speed limit $M_2$*
Upper velocity limit for *positive* potential structures arise from the physical requirement of real ion number density, as given by equation (17). So we find the upper possible value of Mach number (say $M_2$) by imposing the requirement $\psi(\phi = \frac{\left(M-\sqrt{3\sigma/Z}\right)^2}{2}) \geq 0$ (the discussion below equation (17) is reminded). The upper Mach number limit cannot be expressed in a simple closed form, but has to be analyzed numerically. In figure 3, we have displayed this requirement numerically to show the variation of upper soliton speed with respect to the ion temperature ($\sigma$). The solid curve represents the Maxwellian limit (high $\kappa$) and the other curves display deviations from the Maxwellian limit via superthermality effect. It is seen that the upper Mach number $M_2$ increases with an increase in ion temperature and also the superthermality lowers the value of $M_2$. Furthermore, our analysis shows that the electron beam density ($\nu$) and the electron beam velocity ($U_0$) don't have considerable effect on the upper soliton speed.

On the other hand, a similar discussion for upper velocity limit for *negative* potential structures arise from the physical requirement of real beam number density, as given by equation (21), in which, the requirement $\psi(\phi = -\frac{Z\mu(M-U_0)^2}{2}) \geq 0$ has to considered. These negative potential solitons will analyse in the following sections.

*6.3. Mach number range: existence regions*
From equation (26), it is seen that both $\psi(\phi = 0, M) = 0$ and $d\psi(\phi, M)/d\phi|_{\phi=0} = 0$ are spontaneously satisfied at equilibrium, so the quasi-neutrality and lack of the electric field at equilibrium state (far from ES excitations) are held. On the other hand, IA solitary waves exist for the values of the Mach number $M$ in the range $M_1 < M < M_2$, which the dependence of lower an upper limits with respect to the different parameters have discussed in details earlier.

In figure 4, the range of permitted values for the soliton speed $M$ (Mach number) is displayed in $\kappa - M$ plane, therein the variation of the lower and upper limit $M_1$ and $M_2$ with $\kappa$ is showed for representative fixed values of the ion temperature $\sigma$, the electron beam density $\nu$ and the beam velocity $U_0$. In this graph, the dottted curve represents the lower limit $M_1$ in the case $\nu = 0$ (no beam), the solid curve displays $M_1$ for $\nu = 0.00011$ and the dashed curve shows the upper limit $M_2$ for $\nu = 0$, or $\nu = 0.00011$ (no considerable dependence of $M_2$ on electron beam parameters (i.e. $\nu$ and $U_0$) is reminded). So, the solitons may occur in the region between the top and middle curves for a warm plasma including electron beam or between the top and bottom curves for a warm *e-i* plasma. An immediate result from this figure is the reduction of Mach number range as the excess of superthermal electrons is increased (i.e. with decreasing $\kappa$), and even tends to zero as $\kappa \to 3/2$. Another important result implies from figure 4, is that the presence of an electron beam in plasma reduces the allowed soliton speed range. In fact, the presence of electron beam increases the lower limit $M_1$, but has no significant effect on upper limit $M_2$ and so the soliton speed range decreases. It is possible, for high values of $\nu$, the lower limit Mach number $M_1$ reach or even exceed the upper limit $M_2$ and thus no solitary wave be possible in the beam-plasma. For instance, we numerically find that for representative values in figure 4, the critical electron-beam density, after which no solitary waves occur is $\nu = 0.00025$.

**7. Solitary waves structure: numerical analysis**
The pseudo-potential function given in equation (26) depends on electrostatic potential $\phi$ and several physical parameters: superthermality index ($\kappa$), ion temperature ($\sigma$), soliton speed ($M$), electron-beam density ($\nu$) and electron-beam velocity ($U_0$). We shall now numerically investigate the effect of each of these parameters on both positive potential pulse and negative potential structures, separately. It is necessary to mention that in pseudo-potential theory the shape of potential pulse is predicted from analysis of pseudo-potential function. The root of pseudo-potential $\psi(\phi)$ (width of potential well) corresponds to the maximum pulse amplitude ($\phi_m$), while the depth of the pseudo-potential is associated with the slope of potential curve $\phi(\xi)$ (see equation (25)). So a deeper potential well implies a steeper (narrower) soliton pulse.

*7.1. superthermality effect (via $\kappa$)*
To show the effect of superthermality, we have considered different values of $\kappa$ and depicted the corresponding pseudo-potential $\psi(\phi)$ in figures 5(a) and 5(b) for both positive and negative potential structures. It is found that the width and depth of potential well increase monotonically as a result of an increase in superthermality (i.e. a decrease in the value of $\kappa$). So, the maximum amplitude of the IA solitons ($\phi_m$) increases and the soliton pulse becomes steeper for a greater excess of superthermal electrons. To confirm the effect of superthermality on potential structures, we have plotted numerically the maximum amplitude $\phi_m$ versus $\kappa$ in figure 5(c) for some allowed values of $M$. We see again that for smaller values of $\kappa$ (more superthermal electrons) the maximum amplitude has larger values relative to high values of $\kappa$ (pseudo-Maxwellian limit). These results agree with the similar results given in Refs. 69 and 70.

Furthermore, inspection of figures 5(a) and 5(b) reveals that the maximum amplitude and width of negative structures are very small, as compared with the positive ones, which agree with our expectation as earlier claimed in the text.

### 7.2. Ion temperature ($\sigma$) effect

Figure 6(a) and 6(b) show the variation of pseudo-potential $\psi(\phi)$ versus $\phi$ for different values of ion temperature ($\sigma$), with $\kappa = 4.5$ and fixed values of other plasma parameters. We see that the maximum amplitude of ES solitons decreases as the ion temperature is increased. It is also found that the depth of the potential well (and hence the steepness of the soliton pulse) decreases as the ion temperature is increased. So, the presence of warm ions in plasma results in propagation of wider solitons with smaller amplitude.

### 7.3. Soliton speed ($M$) effect

Figure 7(a) and 7(b) have depicted the variation of the Sagdeev pseudo-potential curves for a set of values of $M$ within the range $M_1 < M < M_2$ with $\kappa = 4.5$ and keeping other plasma parameters fixed. It can be seen from these figures that for both type of positive potential wells and negative ones, increasing the soliton speed will increase the width and depth of potential well. So, the amplitude and steepness of soliton profile increase with an increase in soliton speed. This aspect are explored further in figure 7(c), where the variation of the maximum amplitude with respect to soliton speed has plotted numerically for different values of $\kappa$. It is clearly revealed from this figure that a faster soliton has a taller amplitude pulse or inversely, the larger excitations propagate at higher speeds. It is also confirmed that an increase in excess superthermal electrons in the plasma leads to propagation of solitons occurs in Mach number domains with slower speeds as compared to the Maxwellian plasmas (trace the smaller values of $\kappa$ in figure 7(c)).

### 7.4. Beam density ($\nu$) effect

Variation of Sagdeev pseudo-potential structures with beam density $\nu$ for regions of positive potential, practically have no noticeable change. But, for negative potential regions we have depicted in figure 8 the variation of potential well as a result of an increase in electron beam density. It is seen that for ES solitons with negative polarity, the width and depth of potential well decrease and results in wider soliton (less steepness) with less amplitude. However, a subtle investigation on structure of ES solitary waves as a result of change in electron-beam density arise from figures 9(a) and 9(b), where we have analyzed numerically dependence of maximum amplitude of potential structures on the beam density for both positive and negative ES excitations. It is found from figures 9(a) that positive potential solitary waves have a small but finite decrease in amplitude as the electron-beam density increases. In fact, a variation of the order of $10^{-3}$ or less than 0.1% in the amplitude is observed, for the given values. On the other hand, from figures 9(b) for negative potential structures, the absolute value of $\phi_m$ decreases with increasing $\nu$. Anyway, since the negative ES solitons have small amplitudes, there is a critical electron beam density, after which these structures disappear. For instance, we find numerically that for the given values in figure 9(b), i.e. $\kappa = 4$, $\sigma = 0.1$, $M = 1.12$ and $U_0 = 0.05$, the critical electron-beam density is obtained as $\nu = 0.00025$.

## 7.5. Beam velocity ($U_0$) effect

For investigation of beam velocity effect on solitary waves, there is a fundamental distinction between co- and counter propagation electron beam with respect to solitary wave propagation direction. Similar to the beam density effect, the potential well variation in positive regions as a result of electron-beam speed has no perceptible change. Figures 10(a) and 10(b) depict effect of electron-beam speed on Sagdeev pseudo-potential in the negative potential regions for co-propagation and counter-propagation of beam flow and ES solitary waves, respectively. It is seen that, in the case of co-propagating solitons, a higher beam speed leads to a decrease in the solitary wave amplitude (in fact, in absolute value) and to an increase in its width (see figure 10(a)). But, in the case of counter-propagating solitary waves, the inverse result is observed, where the soliton amplitude and its steepness increase for higher beam velocities. (see figure 10(b)). Furthermore, an inspection of analysis obtained here, exhibit the higher soliton amplitude $\phi_m$ in the case of counter-propagation of soliton and beam as compared with the co-propagation case, which agrees with similar results of Ref. 42. Additionally, we have analyzed numerically the variation of the maximum amplitude of both positive and negative ES excitations versus the beam speed for solitary waves propagating either in the direction of the beam flow or in opposite direction to it, as shown in figure 11. Considering the positive potential pulses, with increasing the beam velocity there is an infinitesimally small change (in fact, a decrease in the co-propagation case and an increase in the counter-propagation case) in pulse amplitude (see figure 11(a)). This small change is of the order of $10^{-5}$ or less than 0.001% for the given values. Analogous consideration for negative potential solitons is depicted in figure 11(b), where the variation of maximum amplitude with the beam velocity has been showed. We see that co-propagating negative ES solitons decrease in amplitude with an increase in the value of beam velocity. In fact, there is an upper beam velocity, after which these negative excitations disappear. Our numerical analysis shows that for the given values in figure 11(b), i.e. $\kappa = 4$, $\sigma = 0.1$, $M = 1.1$ and $\nu = 0.00011$, when the beam velocity exceeds the critical value $U_0 = 0.12$, no negative ES soliton is possible. On the other hand, for such negative pulses the amplitude increases for higher beam velocity in the counter-propagating case.

Discussions on two later subsections reveals an important result, arise from the interaction of IASWs with the electron beam in plasma. We see in the counter-propagation of soliton and beam, an increasing in the beam speed $U_0$ results in growing the maximum amplitude $\phi_m$ yet, in the co-propagation case, the result is reversed. Our postulation is that in the counter-propagation case, there is a higher interaction between IASWs and electron beam, in comparison with co-propagation case. Since, in the counter-propagation case, the beam current $j = \nu U_0$ is in opposite direction relative to the propagation of IASWs and so, increasing the beam speed excites this interaction. On the other hand, in the co-propagation case, the beam current $j = \nu U_0$ is in direction of propagation of IASWs and hence, increasing the beam speed lowers this interaction. Anyway, higher soliton amplitude $\phi_m$ in the case of counter-propagating solitons with respect to beam, as compared with the co-propagating case, confirms this claim (see figures 10(a) and 10(b)).

## 7.6. Small amplitude soitons

In section 5, we showed the small-amplitude limit for ES excitations, by Taylor expanding pseudo-potential $\psi(\phi, M)$ near $\phi = 0$, and obtained a solitary wave solution given in equation (30). In figure 12, we have plotted the associated potential $\phi(\xi)$ solitons for different values of $\kappa$, from strongly superthermal electrons (small values of $\kappa$) up to pseudo-Maxwellian ones (high values of $\kappa$). We see that the superthermality increase the soliton amplitude and its steepness, as discussed in details earlier. Furthermore, analysis shows that the small-amplitude ES solitons happen only in the form of *negative* potential excitations, regardless of every choice for allowed values of other physical parameters. This observation has not to be surprising, since the formation of Sagdeev's pseudo-potential with small values is solely restricted to the negative potential regions, as we predicted it both in deriving energy-balance equation and in numerical analysis of pseudo-potential.

**8. Conclusions**

In this paper, we have studied the linear and nonlinear properties of IASWs propagating in a three component unmagnetized plasma consisting of a warn ion fluid, superthermal electrons (whose distribution is modeled by a kappa distribution) and an electron beam. In the linear regime, the dispersion relation for IA waves has obtained by linearizing of basic equations. On the other hand, in nonlinear analysis, a Sagdeev's pseudo-potential approach has been used to obtain the energy integral and the energy-balance like equation which describes the dynamics of IA solitary waves. The coexistence of compressive and rarefactive solitary waves in presence of an electron drifting beam is studied and effects of superthermality, ion temperature, soliton speed and electron-beam parameters (density and velocity) on criteria which support the allowed Mach number range, have been parametrically investigated. Furthermore, effects of mentioned parameters on characteristics of IA solitons have been numerically analyzed.

The results are summarized as follows:

(i) In the linear dispersion relation, increasing the superthermality lowers the phase speed $v_{ph} = \frac{\omega}{k}$ for IA waves, suggesting that IA modes have an upper boundary phase speed at Maxwellian limit.

(ii) It is found that the inclusion of superthemal particles, thermal ions and electron beam significantly modifies the characteristics of IA solitons. The possibility of both positive and negative polarity solitons are predicted in such plasma. Further, it is observed that negative structures are very small, as compared with the positive ones.

(iii) We have shown that the soliton speed threshold ($M_1$) decreases with an increase in the superthermality (i.e. with decreasing $\kappa$). On the other hand, the presence of thermal ions (represented via $\sigma$) and the electron beam (confirmed by $\nu$) cause the soliton speed threshold to increase. Furthermore, in different parametrical limits for soliton speed threshold, the expected results associated with the IA solitons in plasma are recovered, e.g. cold-ion plasma limit ($\sigma \to 0$), Maxwellian plasma limit ($\kappa \to \infty$), plasma with no beam ($\nu \to 0$) and the mixtures of these limits. Additionally, it is observed that the solitons found for either subsonic or supersonic speeds, depend on the value of ion temperature ($\sigma$) and measure of superthermality for electrons in the plasma.

(iv) It is found that the upper soliton speed limit ($M_2$) decreases with an increased excess superthermal electrons, but $M_2$ increases as a result of higher temperatures for ion. However, the electron-beam parameters, i.e. $\nu$ and $U_0$, do not have a significant qualitative effect on the upper soliton speed limit.

(v) It is revealed that the Mach number range, $M_1 < M < M_2$, strongly depends on the suprthermality and electron beam parameters. It is observed that the range of accessible Mach number decreases as the excess of superthermal electrons is increased (i.e. with decreasing $\kappa$) and this range tends to zero as $\kappa \to 3/2$. Furthermore, the presence of electron-beam lessens the Mach number range. We have showed that there are upper critical values for beam parameters (i.e. for density $\nu$ and velocity $U_0$) after which, no ES solitary wave is possible to propagate in plasma.

(vi) We have showed that solitons in plasmas with lower values of $\kappa$ have larger amplitude and steeper profile than the cases with high values of $\kappa$ (pseudo-Maxwellian limit). It is also confirmed that an increased excess superthermal electrons in the plasma leads to the propagation of solitons occurs in Mach number domains with slower speeds as compared to the Maxwellian plasmas. On the other hand, increasing the ion temperature ($\sigma$) yields a decrease in soliton amplitude and less pronounced steepness of soliton profile. It is also found that soliton amplitude and soliton profile steepness both increase monotonically as the soliton speed ($M$) is increased. Furthermore, both of the amplitude and the profile steepness of solitons decrease monotonically with an increase in the value of electron beam density ($\nu$). Specially, in the case of negative structures, it is observed that after a critical beam density the solitons are disappeared. It is also observed that the dependency of positive potential structures on beam parameters is weak.

(vii) There is a fundamental distinction between co- and counter- propagating solitary waves with respect to the electron-beam flow, arise from a typical interaction between IASWs and the electron beam in plasma. We have shown that in the counter-propagating case, an increasing in the beam speed $U_0$ results in growing the maximum amplitude $\phi_m$ yet, in the co-propagating case the result is reversed, where the co-propagating ES solitons decrease in amplitude with an increase in the absolute value of the beam velocity. Furthermore, higher soliton amplitude $\phi_m$ in the counter-propagating case is seen, as compared with the co-propagating case.

(viii) Finally, our analysis has shown that the small-amplitude ES solitons happen only in the form of *negative* potential excitations, regardless of every choice for allowed values of other physical parameters.

Our results may be useful in understanding the various aspects of ES excitations which propagate in space plasmas, where the presence of superthermal are very common, as well as in laboratory experiments, where it is may an acceleration mechanism leads to non-Maxwellian distribution in plasma. In such plasmas, particularly in space plasmas, the presence of an electron drifting beam has frequently reported and also ionic thermal effects has not to be ignored.

**References**

[1] Fried B D and Gould R W 1961 *Phys. Fluids* **4** 139
[2] Yu M Y and Shukla P K 1983 *J. Plasma Phys.* **29** 409
[3] Watanabe K and Taniuti T 1977 *J. Phys. Soc. Jpn.* **43** 1819
[4] Nakamura Y, Ito T and Koga K 1993 *J. Plasma Phys.* **49** 331
[5] Pottelette R, Ergun R E, Treumann R A, Berthomier M, Carlson C W, McFadden J P and Roth I 1999 *Geophys. Res. Lett* **26** 2629
[6] Shukla P K 2003 *Phys. Plasmas* **10** 1619
[7] Xiao D L, Ma J X, Li Y F, Xia Y H and Yu M Y 2006 *Phys. Plasmas* **13** 052308
[8] Misra A P and Bhowmik C 2007 *Phys. Lett. A* **369** 90
[9] Boström R, Gustafsson G, Holback B, Holmgren G and Koskinen H 1988 *Phys. Rev. Lett.* **61** 82
[10] Ergun R E, Carlson C W, McFadden J P, Mozer F S, Delory G T, Peria W, Chaston C C, Temerin M, Roth I, Muschietti L, Elphic R, Strangeway R, Pfaff R, Cattell C A, Klumpar D, Shelley E, Peterson W, Moebius E and Kistler L 1998 *Geophys. Rev. Lett.* **25** 2041
[11] Ergun R E, Andersson L, Tao J, Angelopoulos V, Bonnell J, McFadden J P, Larson D E, Eriksson S, Johansson T, Cully C M, Newman D N, Goldman M V, Roux A, Lecontel O, Glassmeier K H and Baumjohann W 2009 *Phys. Rev. Lett.* **102** 155002
[12] Popel S I, Vladimirov S V and Shukla P K 1995 *Phys. Plasmas* **2** 716
[13] Nejoh Y N 1996 *Phys. Plasma* **3** 1447
[14] Nejoh Y N 1997 *Australian J. Phys*. **50** 309
[15] Yu M Y, Shukla P K and Bujarbarua S 1980 *Phys. Fluids* **23** 2146
[16] Hasegawa H, Irie S, Usami S and Ohsawa Y 2002 *Phys. Plasmas* **9** 2549
[17] Mahmood S, Mushtaq A and Saleem H 2003 *New J. Phys.* **5** 28
[18] Saberian E, Esfandyari-Kalehjahi A R and Akbari-Moghanjoughi 20111 *Can. J. Phys.* **89** 299
[19] Sahu B 2012 *Astrophys. Space Sci.* **338** 251
[20] Sagdeev R Z 1966 *Rev. Plasma Phys.* edited by Leontovich M A Consultants Bureau **4** 23
[21] Ergun R E *et al* 1998 *Geophys. Res. Lett.* **25** 2061
    Delory G T *et al* 1998 *Geophys. Res. Lett.* **25** 2069
    Pottelette R *et al* 1999 *Geophys. Res. Lett.* **26** 2629
[22] McFadden J P *et al* 2003 *J. Geophys. Res.* **108** 8018
[23] Temerin M 1982 *Phys. Rev. Lett.* **48** 1175
[24] Boström R 1988 *Phys. Rev. Lett.* **61** 82
[25] Matsumoto H *et al* 1994 *Geophys. Res. Lett.* **21** 2915
    Franz J R *et al* 1998 *Geophys. Res. Lett.* **25** 1277
    Cattell C A *et al* 1999 *Geophys. Res. Lett.* **26** 425
[26] Yajima N, Tanaka M, Itoh T, Nakayama T, Takeda T and Kamiya T 1996 *Proc. 1996 ICPP (Nagoya)* **1** 774.
[27] Yamagiwa K, Itoh T and Nakayama T 1997 *J. Phys. IV* **7** C4
[28] Takeda T and Yamagiwa K 2003 *J. Plasma Fusion Res.* **79** 323
[29] Chatterjee P, Roychoudhury R, Naturforsch Z A 1995 *Phys. Sci.* **51** 1002
[30] Nejoh Y and Sanuki H 1995 *Phys. Plasmas* **2** 4122
[31] Nejoh Y 1997 *J. Plasma Phys.* **56** 67
    Nejoh Y 1997 *J. Plasma Phys.* **57** 841
[32] Moslem W M 1999 *J. Plasma Phys.* **61** 177
[33] Moslem W M 2000 *J. Plasma Phys.* **63** 139
[34] Esfandyari A, Khorram S and Rostami A 2001 *Phys. Plasmas* **8** 4753
[35] Berthomier M, Pottelette R, Malingre M and Khotyaintsev Y 2000 *Phys. Plasmas* **7** 2987
[36] Lu Q and Wang S 2005 *Proceedings of ISSS-7* Kyoto
[37] Gell Y and Roth T 1977 *Plasma Phys.* **19** 915
[38] Sahu B and Roychoudhury R 2004 *Phys. Plasmas* **11** 1947
[39] El-Taibany W F and Moslem W M 2005 *Phys. Plasmas* **12** 032307
[40] Lakhina G S, Singh S V, Kakad A P, Verheest F and Bharuthram R 2008 *Nonlinear Process. Geophys.* **15** 903

[41] Esfandyari-Kalejahi A, Kourakis I and Shukla P K 2008 *Phys. Plasmas* **15** 022303
[42] Devanandhan S, Singh S V, Lakhina G S and Bharuthram R 2011 *Nonlin. Processes Geophys.* **18** 627
[43] Montgomery M D, Bame S J and Hundhause A J 1968 *Geophys. J. Res.* **73** 4999
[44] Feldman W C, Asbridge J R, Bame S J, Montgomery M D and Gary S P 1975 *J. Geophys. Res.* **80** 4181
[45] Pilipp W G, Miggenrieder H, Montgomery M D, Muhlhauser K H, Rosenbauer H and Schwenn R 1987 *J. Geophys. Res.* **92** 1075
[46] Maksimovic M, Pierrard V and Riley P 1997 *Geophys. Res. Let.* **24** 1151
[47] Zouganelis I 2008 *J. Geophys. Res.* **113** A08111
[48] Vasyliunas V M 1968 *J. Geophys. Res.* **73** 2839
[49] Gloeckler G, Geiss J, Balsiger H, Bedini P, Cain J C et al 1992 *Astron. Astrophys. Suppl. Ser.* **92** 267
[50] Maksimovic M, Pierrard V and Riley P 1997 *Geophys. Res. Let.* **24** 1151
[51] Retherford K D, Moos H W, Strobel D F 2003 *J. Geophys. Res.* **108** 1333
[52] Mauk B.H et al 2004 *J. Geophys. Res.* **109** A09S12
[53] De la Haye V, Waite J J H, Johnson R E, Telle RV, Cravens T E, Luhmann J G, Kasprzak W T, Gell D A, Magee B, Leblanc F, Michael M, Jurac S and Robertson I P 2007 *J. Geophys. Res.* **112** A07309
[54] Schippers P et al 2008 *J. Geophys. Res.* **113** A07208
[55] Dialynas K, Krimigis SM, Mitchemm D G, Hamilton D C, Krupp N and Brandt P C 2009 *J. Geophys. Res.* **114** A01212
[56] Hasegawa A, Mima K and Duong-van M 1985 *Phys. Rev. Lett.* **54** 2608
[57] Collier M C 1993 *Geophys. Res. Lett.* **20** 1531
[58] Treumann R A 2001 *Astrophys. Space Sci.* **277** 81
[59] Leubner M P 2002 *Astrophys. Space Sci.* **282** 573
[60] Tsallis C 1995 *Phys. A* **221** 277
[61] Leubner M P and Voros Z 2005 *Astrophys. J.* **618** 547
[62] Treumann R A and Jaroschek C H 2008 *Phys. Rev. Lett.* **100** 155005
[63] Leubner M P 2008 *Nonlin. Proc. Geophys.* **15** 531
    Leubner M P and Schupfer N 2000 *J. Geophys. Res.* **105b** 27387
[64] Livadiotis G and McComas D J 2009 *J. Geophys. Res.* **114** A11105
[65] Vocks C and Mann G 2003 *Astroph. J.* **593** 1134
[66] Vocks C, Mann G and Rausche G 2008 *Astron. Astrophys.* **480** 527
[67] Pierrard V and Lazar M 2010 *Solar Phys.* **267** 153
[68] Chen H and Liu S Q 2012 *Astrophys. Space Sci.* **339** 179
[69] Saini N S, Kourakis I and Hellberg M A 2009 *Phys. Plasmas* **16** 062903
[70] Sultana S, Kourakis I, Saini N S and Hellberg A 2010 *Phys. Plasmas* **17** 032310
[72] Saini N S and Kiurakis I 2010 *Plasma Phys. Control. Fusion* **52** 075009
[73] Gogoi R, Roychoudhary R and Khan M 2011 *Indian Journal of Pure & Applied Physics* **49** 173
[74] Mamun A A 1997 *Phys. Rev. E* **55** No. 2 1852
[76] Ju-Kue X *et al* 2002 *Chin. Phys. Soc.* **11** No. 11 1184
[77] Jukui X 2003 *Chaos, Solitons and Fractals* **18** 849
[78] Verheest F, Helberg M A and Lakhina G S 2007 *Astrophys. Space Sci. Trans.* **3** 15
[79] Mahmood S and Akhtar N 2008 *Eur. Phys. J. D* **49** 217
[80] Chattopadhyay S 2010 *FIZIKA A* **19** 31
[81] Berthomier M, Pottelette R, Malinger M and Khotyaintsev Y 2000 *Phys. Plasmas* **7** 2987
[82] Treumann R A and Baumjohann W 1997 *"Advanced Space Plasma Physics "* Imperial Colleage Press London p. 260

**Figures caption:**

**Figure 1.** Variation of the linear dispersion relation for propagating IA modes for various values of $\kappa$, from high superthemal electrons (small values of $\kappa$) to pseudo-Maxwellian ones (high values of $\kappa$) for $Z = 1$, $\sigma = 0.1$, $\nu = 0.00018$ and $U_0 = 0.05$.

**Figure 2.** Variation of the threshold soliton speed $M_1$ (a) versus ion temperature ($\sigma$) for $\nu = 0.00025$ and $U_0 = 0.05$, (b) versus electron-beam density ($\nu$) for $\sigma = 0.1$ and $U_0 = 0.05$ and (c) versus electron-beam velocity ($U_0$) for $\sigma = 0.1$ and $\nu = 0.00025$. From bottom to top, the dotted curve corresponds to $\kappa = 2$, dashed to $\kappa = 3$, dotted-dashed to $\kappa = 5$ and solid curve to $\kappa = 16$.

**Figure 3.** Variation of upper soliton speed limit $M_2$ versus ion temperature ($\sigma$) for $\nu = 0.00025$ and $U_0 = 0.05$. From bottom to top, the dotted curve correspond to $\kappa = 2$, dashed to $\kappa = 3$, dotted-dashed to $\kappa = 5$ and solid curve to $\kappa = 16$.

**Figure 4.** The allowed Mach number ($M$) range for existence of IA solitary waves in the $\kappa - M$ plane with $\sigma = 0.1$ and $U_0 = 0.05$. From bottom to top, the dotted curve represents the lower limit $M_1$ in the case $\nu = 0$ (no beam), the solid curve displays $M_1$ for $\nu = 0.00011$ and the dashed curve shows the upper limit $M_2$ for $\nu = 0$, or $\nu = 0.00011$ (the dependence of $M_2$ on electron beam parameters (i.e. $\nu$ and $U_0$) isn't considerable. Solitons may occur in the region between the top and middle curves for a warm plasma including electron beam or between the top and bottom curves for a warm *e-i* plasma.

**Figure 5.** Variation of pseudo-potential $\psi(\phi)$ versus $\phi$ for $\sigma = 0.1$, $\nu = 0.00011$, $U_0 = 0.05$ and different values of $\kappa$ (a) in positive potential regions with $M = 1.14$ and (b) in negative potential regions with $M = 1.25$. From bottom to top, the dotted-dashed curve corresponds to $\kappa = 4.7$, dashed to $\kappa = 6$ and solid curve to $\kappa = 8$. (c) Variation of the maximum amplitude of the IA solitons ($\phi_m$) versus $\kappa$ for $\sigma = 0.1$, $\nu = 0.00011$, $U_0 = 0.05$ and some allowed values of $M$.

**Figure 6.** Variation of pseudo-potential $\psi(\phi)$ versus $\phi$ for $\kappa = 4.5$, $\nu = 0.00011$, $U_0 = 0.05$ and various temperatures for ions (a) in positive potential regions with $M = 1.14$. From top to bottom, the dotted-dashed curve: $\sigma = 0.11$; dashed curve: $\sigma = 0.13$; and solid curve: $\sigma = 0.15$ and (b) in negative potential regions with $M = 1.25$. From top to bottom, the dotted- dashed curve: $\sigma = 0.12$; dashed curve: $\sigma = 0.18$; and solid curve: $\sigma = 0.20$

**Figure 7.** Variation of pseudo-potential $\psi(\phi)$ versus $\phi$ for different soliton speeds $M$ and $\kappa = 4.5$, $\sigma = 0.1$, $\nu = 0.00018$ and $U_0 = 0.05$ (a) in positive potential regions. From top to bottom, the solid curve: $M = 1.11$; dashed curve: $M = 1.12$; and dotted-dashed curve $M = 1.13$ and (b) in negative potential regions. From top to bottom, the solid curve: $M = 1.25$; dashed curve: $M = 1.27$; and dotted-dashed curve $M = 1.29$. (c) Variation of the maximum amplitude of the IA solitons ($\phi_m$) versus $M$ for $\sigma = 0.1$, $\nu = 0.00018$, $U_0 = 0.05$ and different values of $\kappa$.

**Figure 8.** Variation of Sagdeev's pseudo-poptential structure $\psi(\phi)$ versus $\phi$ for regions of negative potential polarity for different electron-beam densities $\nu$ and $\kappa = 4$, $\sigma = 0.1$, $M = 1.25$ and $U_0 = 0.05$. From bottom to top, the solid curve corresponds to $\nu = 0.00015$, dashed to $\nu = 0.00020$ and dotted-dashed curve to $\nu = 0.00025$. (No noticeable change in regions of positive potential polarity on $\nu$).

**Figure 9**. Variation of the maximum amplitude of the IA solitons ($\phi_m$) versus electron-beam density $\nu$ for $\kappa = 4$, $\sigma = 0.1$, $M = 1.12$ and $U_0 = 0.05$ (a) for positive IA structures and (b) for negative IA structures.

**Figure 10.** Variation of pseudo-potential $\psi(\phi)$ versus $\phi$ in negative potential regions for different values of the electron-beam velocity $U_0$ and $\kappa = 4$, $\sigma = 0.1$, $M = 1.25$ and $\nu = 0.00018$ (a) in the case of co-propagating soliton with electron beam. Solid curve: $U_0 = 0.05$; dashed curve: $U_0 = 0.09$; and dotted-dashed curve $U_0 = 0.12$ and (b) in the case of counter-propagating soliton with electron beam. Solid

curve: $U_0 = -0.05$; dashed curve: $U_0 = -0.09$; and dotted-dashed curve: $U_0 = -0.12$. (No noticeable change in regions of positive potential polarity on $U_0$).

**Figure 11.** Variation of the maximum amplitude of the IA solitons ($\phi_m$) versus electron-beam velocity $U_0$ for $\kappa = 4$, $\sigma = 0.1$, $M = 1.1$ and $\nu = 0.00011$ (a) for positive IA excitations and (b) for negative IA excitations.

**Figure 12.** Typical small-amplitude IA potentials $\phi$ versus $\xi$ for different values of $\kappa$ and $\sigma = 0.1$, $M = 1.25$, $\nu = 0.00011$ and $U_0 = 0.05$. The small-amplitude ES solitons happen only in the form of negative potential excitations.

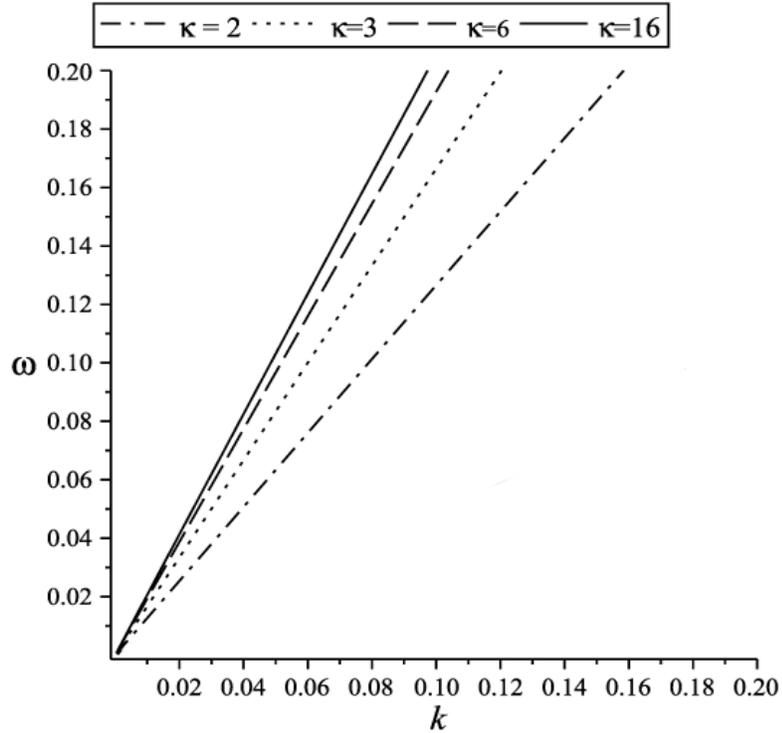

**Figure 1.** Variation of the linear dispersion relation for propagating IA modes for various values of $\kappa$, from high superthemal electrons (small values of $\kappa$) to pseudo-Maxwellian ones (high values of $\kappa$) for $Z = 1$, $\sigma = 0.1$, $\nu = 0.00018$ and $U_0 = 0.05$.

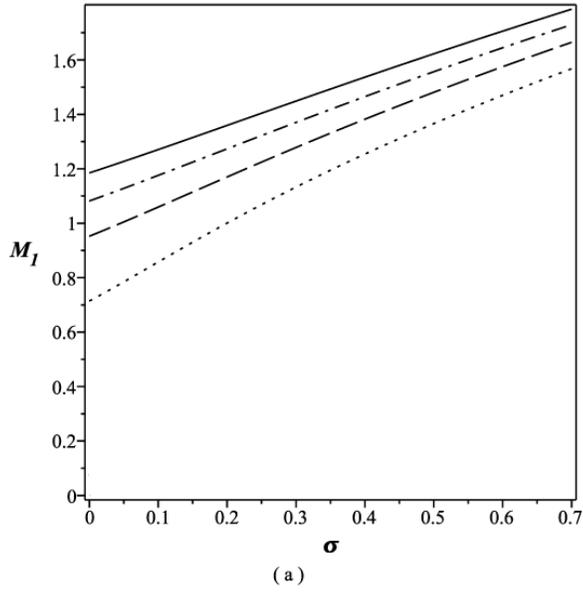
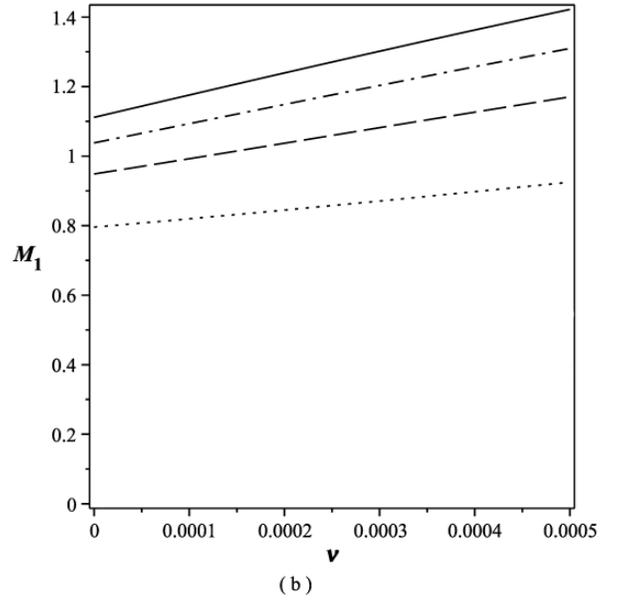
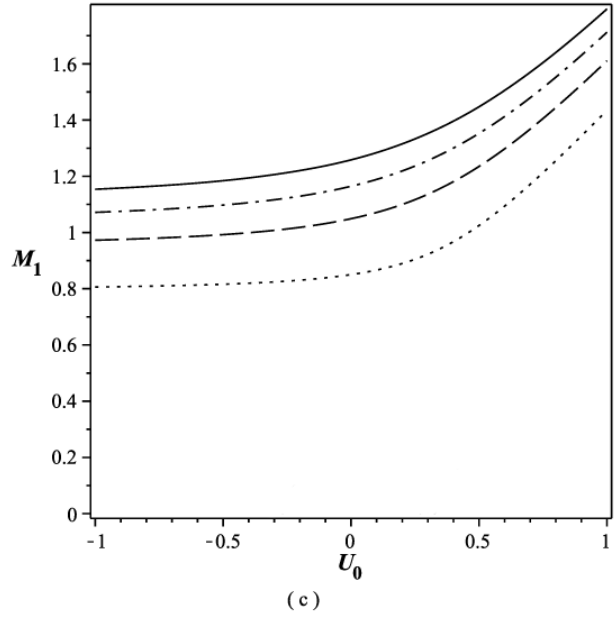

**Figure 2.** Variation of the threshold soliton speed $M_1$ (a) versus ion temperature ($\sigma$) for $\nu = 0.00025$ and $U_0 = 0.05$, (b) versus electron-beam density ($\nu$) for $\sigma = 0.1$ and $U_0 = 0.05$ and (c) versus electron-beam velocity ($U_0$) for $\sigma = 0.1$ and $\nu = 0.00025$. From bottom to top, the dotted curve corresponds to $\kappa = 2$, dashed to $\kappa = 3$, dotted-dashed to $\kappa = 5$ and solid curve to $\kappa = 16$.

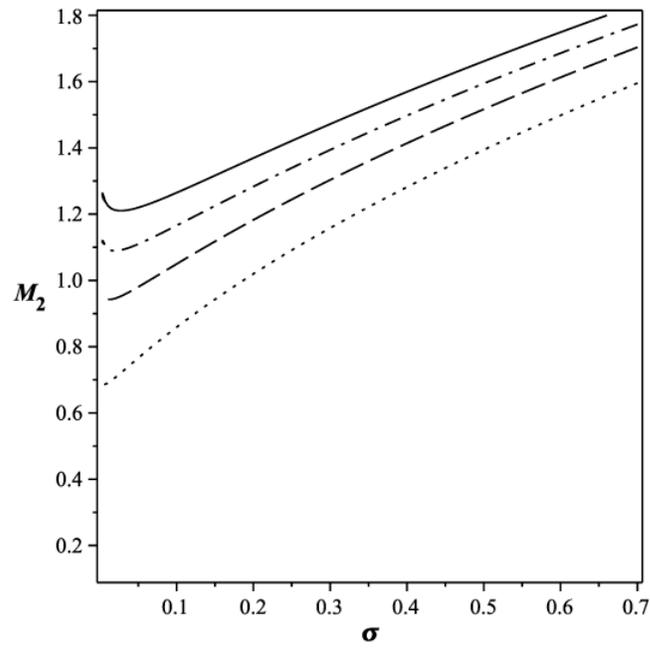

**Figure 3.** Variation of upper soliton speed limit $M_2$ versus ion temperature ($\sigma$) for $\nu = 0.00025$ and $U_0 = 0.05$. From bottom to top, the dotted curve correspond to $\kappa = 2$, dashed to $\kappa = 3$, dotted-dashed to $\kappa = 5$ and solid curve to $\kappa = 16$.

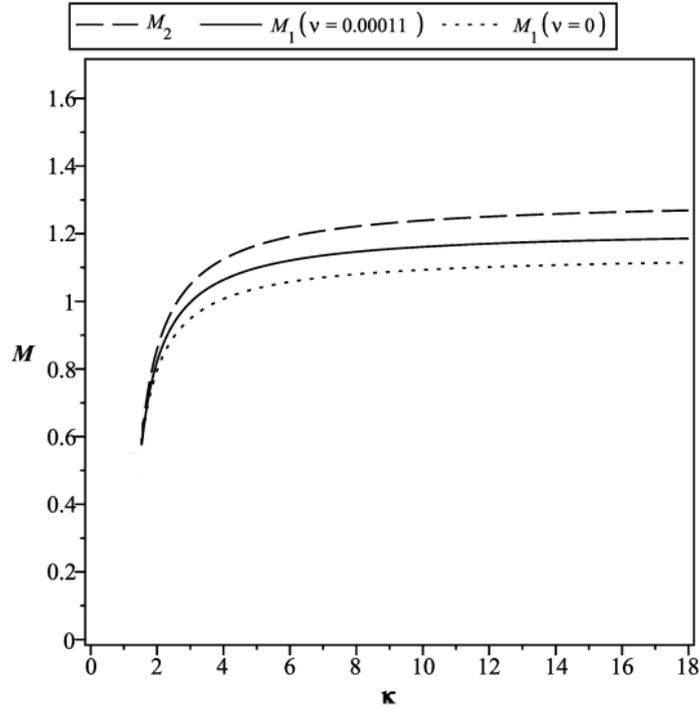

**Figure 4.** The allowed Mach number ($M$) range for existence of IA solitary waves in the $\kappa - M$ plane with $\sigma = 0.1$ and $U_0 = 0.05$. From bottom to top, the dotted curve represents the lower limit $M_1$ in the case $\nu = 0$ (no beam), the solid curve displays $M_1$ for $\nu = 0.00011$ and the dashed curve shows the upper limit $M_2$ for $\nu = 0$, or $\nu = 0.00011$ (the dependence of $M_2$ on electron beam parameters (i.e. $\nu$ and $U_0$) isn't considerable. Solitons may occur in the region between the top and middle curves for a warm plasma including electron beam or between the top and bottom curves for a warm *e-i* plasma.

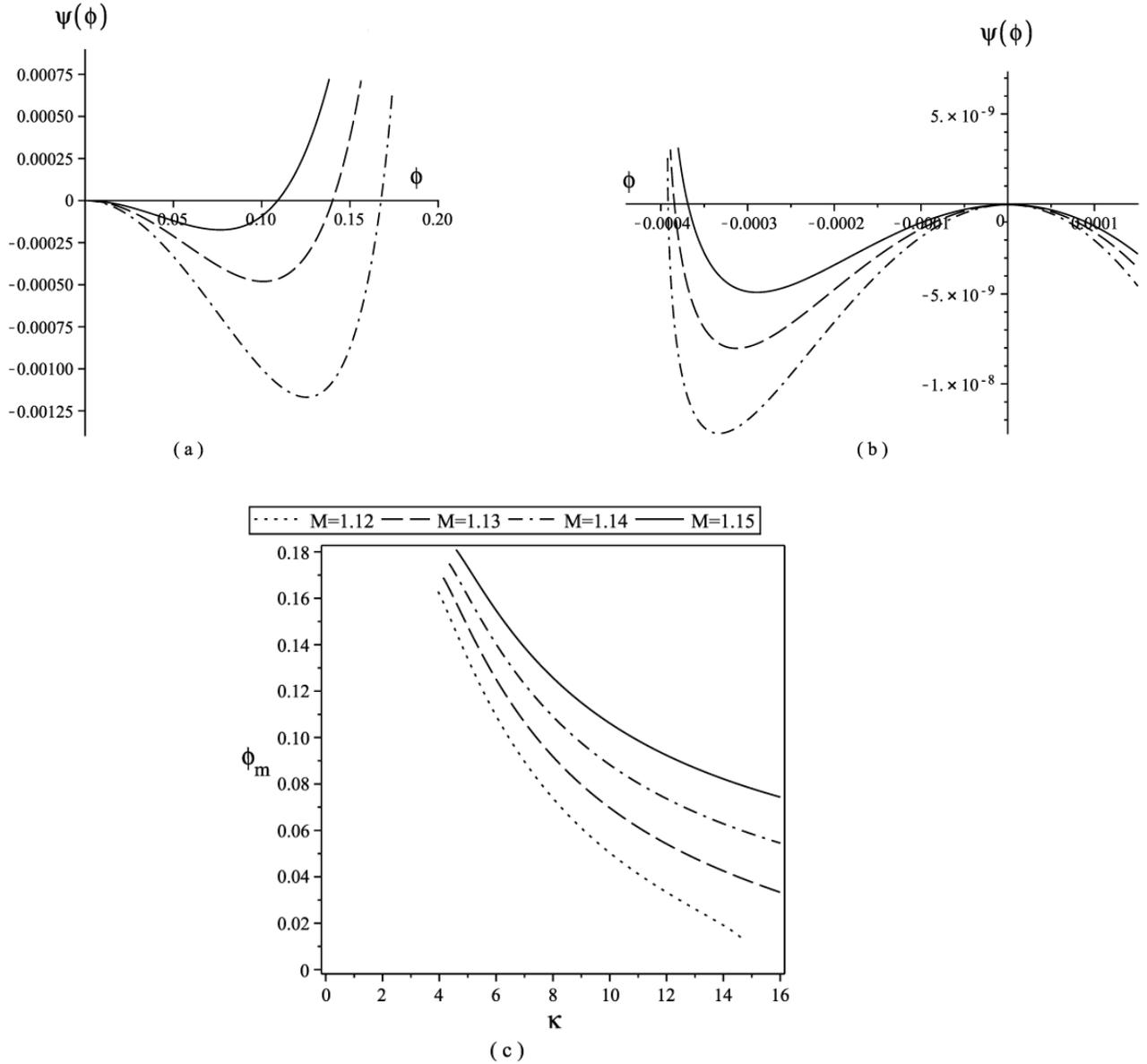

**Figure 5.** Variation of pseudo-potential $\psi(\phi)$ versus $\phi$ for $\sigma = 0.1$, $\nu = 0.00011$, $U_0 = 0.05$ and different values of $\kappa$ (a) in positive potential regions with $M = 1.14$ and (b) in negative potential regions with $M = 1.25$. From bottom to top, the dotted-dashed curve corresponds to $\kappa = 4.7$, dashed to $\kappa = 6$ and solid curve to $\kappa = 8$. (c) Variation of the maximum amplitude of the IA solitons ($\phi_m$) versus $\kappa$ for $\sigma = 0.1$, $\nu = 0.00011$, $U_0 = 0.05$ and some allowed values of $M$.

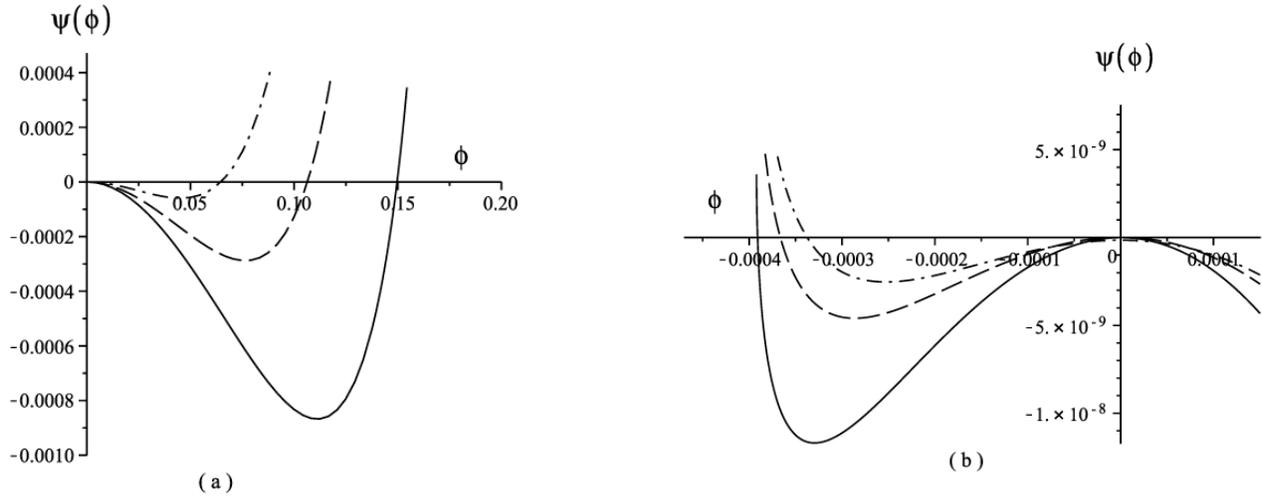

**Figure 6.** Variation of pseudo-potential $\psi(\phi)$ versus $\phi$ for $\kappa = 4.5$, $\nu = 0.00011$, $U_0 = 0.05$ and various temperatures for ions (a) in positive potential regions with $M = 1.14$. From top to bottom, the dotted-dashed curve: $\sigma = 0.11$; dashed curve: $\sigma = 0.13$; and solid curve: $\sigma = 0.15$ and (b) in negative potential regions with $M = 1.25$. From top to bottom, the dotted- dashed curve: $\sigma = 0.12$; dashed curve: $\sigma = 0.18$; and solid curve: $\sigma = 0.20$.

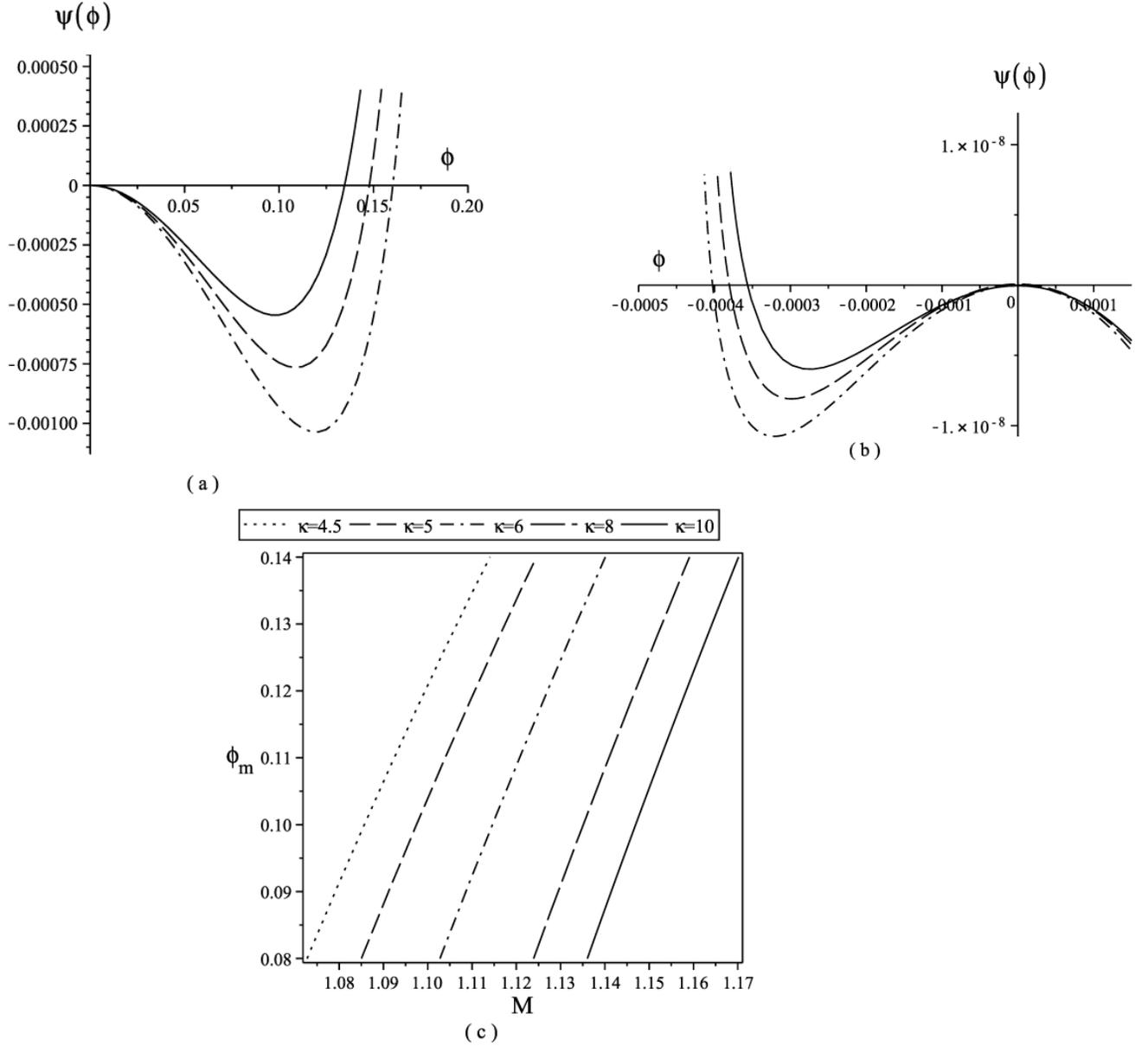

**Figure 7**. Variation of pseudo-potential $\psi(\phi)$ versus $\phi$ for different soliton speeds $M$ and $\kappa = 4.5$, $\sigma = 0.1$, $\nu = 0.00018$ and $U_0 = 0.05$ (a) in positive potential regions. From top to bottom, the solid curve: $M = 1.11$; dashed curve: $M = 1.12$; and dotted-dashed curve $M = 1.13$ and (b) in negative potential regions. From top to bottom, the solid curve: $M = 1.25$; dashed curve: $M = 1.27$; and dotted-dashed curve $M = 1.29$. (c) Variation of the maximum amplitude of the IA solitons ($\phi_m$) versus $M$ for $\sigma = 0.1$, $\nu = 0.00018$, $U_0 = 0.05$ and different values of $\kappa$.

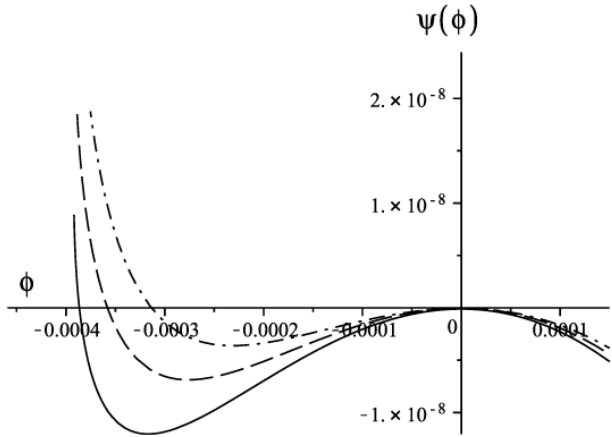

**Figure 8.** Variation of Sagdeev's pseudo-poptential structure $\psi(\phi)$ versus $\phi$ for regions of negative potential polarity for different electron-beam densities $\nu$ and $\kappa = 4$, $\sigma = 0.1$, $M = 1.25$ and $U_0 = 0.05$. From bottom to top, the solid curve corresponds to $\nu = 0.00015$, dashed to $\nu = 0.00020$ and dotted-dashed curve to $\nu = 0.00025$. (No noticeable change in regions of positive potential polarity on $\nu$).

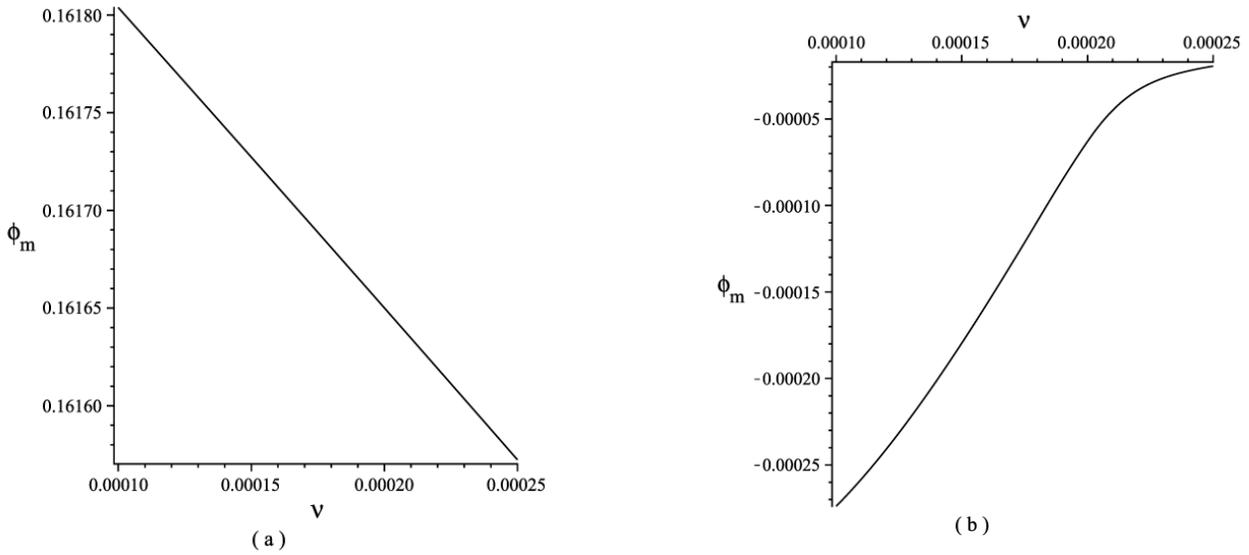

**Figure 9.** Variation of the maximum amplitude of the IA solitons ($\phi_m$) versus electron-beam density $\nu$ for $\kappa = 4$, $\sigma = 0.1$, $M = 1.12$ and $U_0 = 0.05$ (a) for positive IA structures and (b) for negative IA structures.

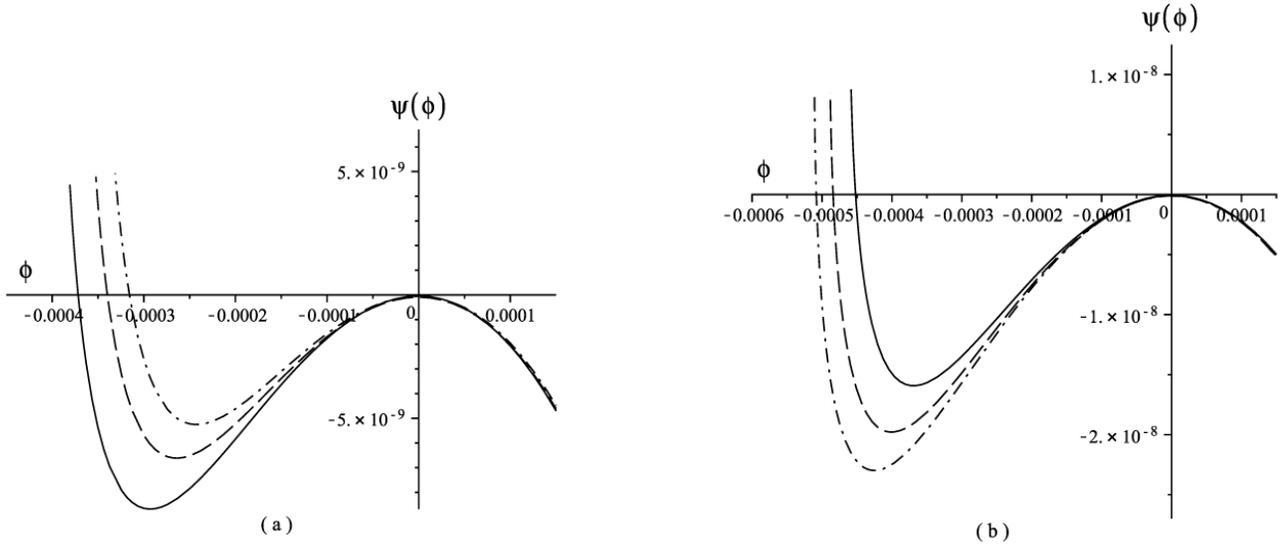

**Figure 10.** Variation of pseudo-potential $\psi(\phi)$ versus $\phi$ in negative potential regions for different values of the electron-beam velocity $U_0$ and $\kappa = 4$, $\sigma = 0.1$, $M = 1.25$ and $\nu = 0.00018$ (a) in the case of co-propagating soliton with electron beam. Solid curve: $U_0 = 0.05$; dashed curve: $U_0 = 0.09$; and dotted-dashed curve $U_0 = 0.12$ and (b) in the case of counter-propagating soliton with electron beam. Solid curve: $U_0 = -0.05$; dashed curve: $U_0 = -0.09$; and dotted-dashed curve: $U_0 = -0.12$. (No noticeable change in regions of positive potential polarity on $U_0$).

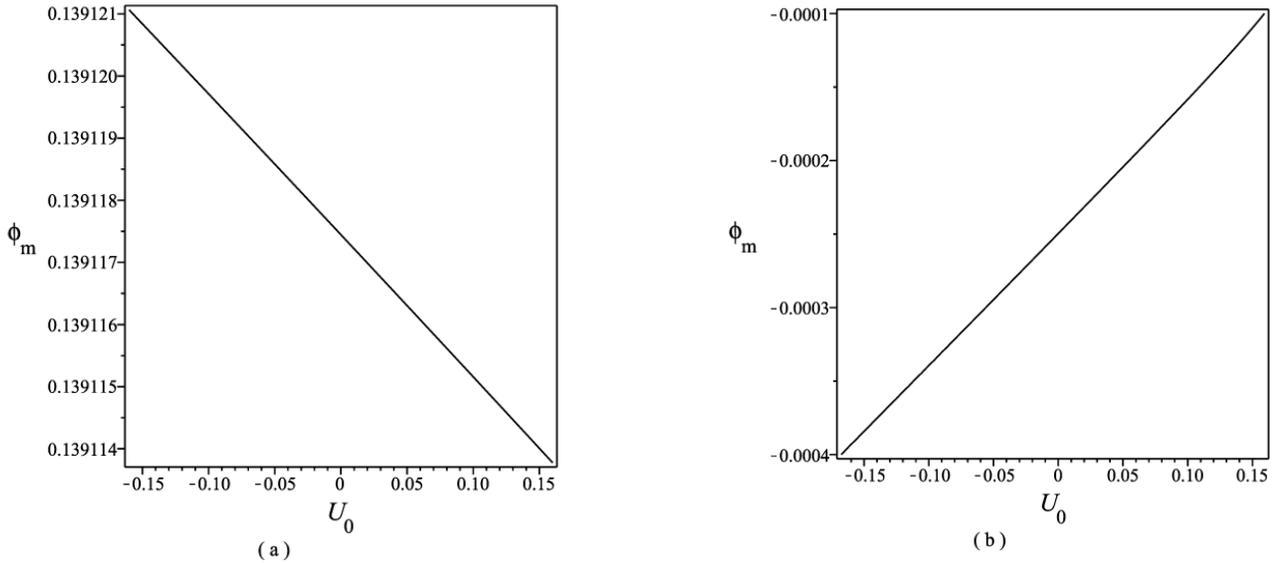

**Figure 11.** Variation of the maximum amplitude of the IA solitons ($\phi_m$) versus electron-beam velocity $U_0$ for $\kappa = 4$, $\sigma = 0.1$, $M = 1.1$ and $\nu = 0.00011$ (a) for positive IA excitations and (b) for negative IA excitations.

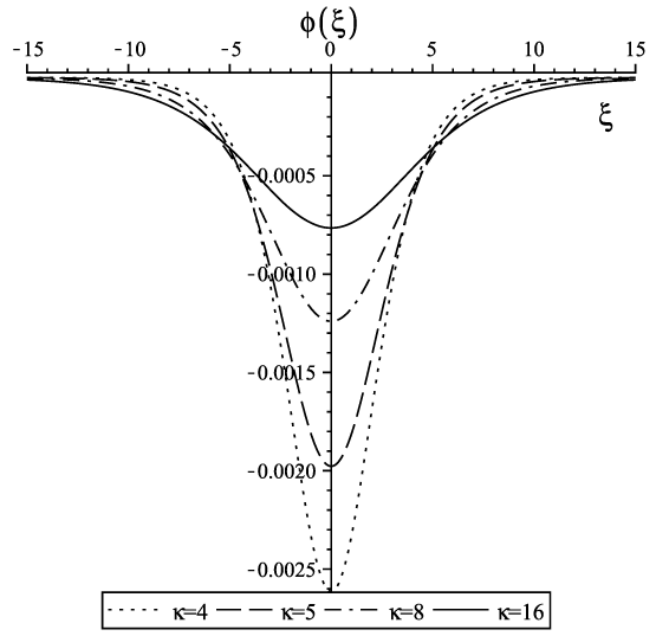

**Figure 12.** Typical small-amplitude IA potentials $\phi$ versus $\xi$ for different values of $\kappa$ and $\sigma = 0.1$, $M = 1.25$, $\nu = 0.00011$ and $U_0 = 0.05$. The small-amplitude ES solitons happen only in the form of negative potential excitations.